\documentclass[prd,twoside,preprintnumbers,superscriptaddress,twocolumn,nofootinbib]{revtex4}

\usepackage{amsmath,graphicx,slashed,multirow,color}

\begin{document}

\preprint{CERN-PH-TH/2014-252}

\title{LHC constraints on gauge boson couplings to dark matter}

\author{Andreas Crivellin}
\email{andreas.crivellin@cern.ch}
\affiliation{CERN Theory Division, CH-1211 Geneva 23, Switzerland}
\author{Ulrich Haisch}
\email{u.haisch1@physics.ox.ac.uk}
\affiliation{Rudolf Peierls Centre for Theoretical Physics, University of Oxford,
OX1 3PN Oxford, United Kingdom}
\affiliation{CERN Theory Division, CH-1211 Geneva 23, Switzerland}
\author{Anthony Hibbs}
\email{Anthony.Hibbs@physics.ox.ac.uk}
\affiliation{Rudolf Peierls Centre for Theoretical Physics, University of Oxford,
OX1 3PN Oxford, United Kingdom}

\date{\today}

\begin{abstract} 
Collider searches for  energetic particles recoiling against missing transverse energy~$(\slashed{E}_T)$ allow to place strong bounds on the interactions between dark matter~(DM) and standard model~(SM) particles. In this article we update and extend LHC constraints on effective dimension-7 operators involving  DM and electroweak gauge bosons. A concise comparison of the sensitivity of the mono-photon, mono-$W$,  mono-$Z$, mono-$W/Z$, invisible Higgs-boson decays in the vector boson fusion~(VBF) mode and the mono-jet channel is presented. Depending on the parameter choices, either the mono-photon or the mono-jet data provide the most stringent bounds at the moment. We furthermore explore the potential of improving the current $8 \, {\rm TeV}$ limits at $14 \, {\rm TeV}$. Future strategies capable of disentangling the effects of the different effective operators involving electroweak gauge bosons are discussed as well. 
\end{abstract}

\maketitle

\section{Introduction}
\label{sec:1}

Because of their potential connection to DM, searches for $\slashed{E}_T$ represent one of the main lines of LHC research. These searches can be categorised based on the type of SM particles that recoil against DM. By now, ATLAS and CMS have considered a plethora of different final states in DM searches containing jets of hadrons, gauge bosons, heavy quarks and even the Higgs boson (see e.g.~\cite{Askew:2014kqa} for a recent review of the experimental status). 

In most cases these studies are performed in the context of an effective field theory (EFT) which correctly captures the physics of heavy particles mediating the interactions between DM and SM fields, if the mediators are heavy enough to be integrated out. Below we will consider the effective Lagrangian 
\begin{equation} \label{eq:1}
{\cal L}_{\rm eff} = \sum_{k=B,W,\tilde B,\tilde W} \frac{C_k (\mu)}{\Lambda^3} \, O_k \,,
\end{equation}
which contains the following four  $SU(2)_L \times U(1)_Y$ gauge-invariant dimension-7 operators
\begin{equation} \label{eq:2}
\begin{split}
O_B= \bar \chi \chi \, B_{\mu \nu} B^{\mu \nu } \,, \qquad 
O_W= \bar \chi \chi \, W_{\mu \nu}^i W^{i, \mu \nu } \,, \\[2mm]
O_{\tilde B}= \bar \chi \chi \, B_{\mu \nu} \tilde B^{\mu \nu } \,, \qquad 
O_{\tilde W}= \bar \chi \chi \, W_{\mu \nu}^i \tilde W^{i, \mu \nu } \,.
\end{split}
\end{equation}
Here $\Lambda$ represents the scale of new physics at which the  higher-dimensional operators~(\ref{eq:1}) are generated, i.e.~the scale where the messenger particles are removed as active degrees of freedom. The DM particle $\chi$ can be both a Dirac or a Majorana fermion and $B_{\mu \nu} = \partial_\mu B_\nu - \partial_\nu B_\mu$ ($W_{\mu \nu}^i =  \partial_\mu W_\nu^i - \partial_\nu W_\mu^i + g_2 \hspace{0.25mm} \epsilon^{ijk}  \hspace{0.25mm}  W_\mu^j \hspace{0.25mm} W_\mu^k$) is the $U(1)_Y$~$\big($$SU(2)_L$$\big)$ field strength tensor, while $\tilde B_{\mu \nu} = 1/2 \hspace{0.5mm} \epsilon_{\mu \nu  \lambda \rho}  \hspace{0.25mm}  B^{\lambda \rho}$ ($\tilde W_{\mu \nu}^i = 1/2 \hspace{0.5mm} \epsilon_{\mu \nu  \lambda \rho}  \hspace{0.25mm}  W^{i, \lambda \rho}$) denotes its dual and $g_2$ is the weak coupling constant.

\begin{figure}[!t]
\begin{center}
\includegraphics[width=0.49 \textwidth]{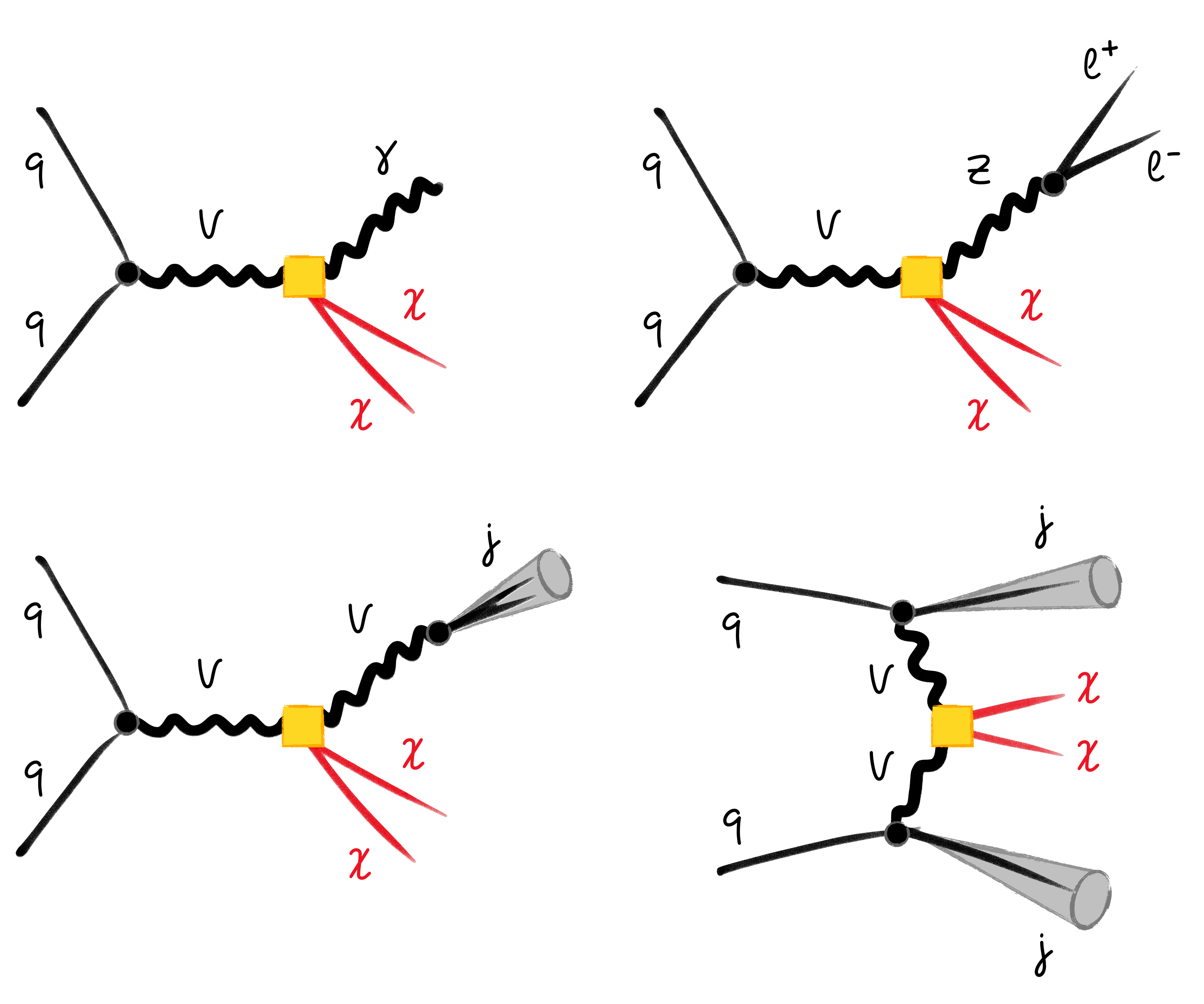} 

\vspace{-2mm}

\caption{\label{fig:1} Representative examples of graphs that generate a $\slashed{E}_T + \gamma$, $\slashed{E}_T + Z \, (\to \ell^+ \ell^-)$, $\slashed{E}_T + W/Z \, (\to j)$ or a  $\slashed{E}_T + 2 j$ signal. The operator insertions are indicated by yellow squares while SM vertices are represented by black dots. Propagators labeled by~$V$ include all possible photon, $Z$-boson or $W$-boson exchanges. See text for further details.}
\end{center}
\end{figure}

The operators introduced in (\ref{eq:2}) appear  in models of Rayleigh DM (see for instance~\cite{Weiner:2012cb,Weiner:2012gm,Liu:2013gba}). They are special in the sense that, up to dimension 7, they are the only effective  interactions which lead to velocity-suppressed annihilation rates of DM to photon pairs~\cite{Rajaraman:2012db,Frandsen:2012db,Rajaraman:2012fu}.  While the sensitivity of future direct detection experiments  may allow to set novel bounds on the Wilson coefficients $C_B (\Lambda)$ and~$C_W (\Lambda)$ for heavy DM particles with $m_\chi \gtrsim 1 \, {\rm TeV}$ once loop effects are taken into account~\cite{Crivellin:2014gpa}, in the case of light DM the leading $\big($and for $C_{\tilde B} (\Lambda)$ and $C_{\tilde W} (\Lambda)$ the only$\big)$ restrictions arise and  will continue to arise from collider searches involving large amounts of~$\slashed{E}_T$. In fact, the DM-SM interactions~(\ref{eq:2}) have been constrained using $7 \, {\rm TeV}$ and $8 \, {\rm TeV}$~LHC data on  invisible decays of the Higgs boson in the VBF mode~\cite{Cotta:2012nj} as well as the $\slashed{E}_T + Z$ \cite{Carpenter:2012rg,Aad:2014vka}, the mono-photon~\cite{Nelson:2013pqa} and the~$\slashed{E}_T + W$~\cite{Lopez:2014qja} channel.  

The main goal of this article is twofold. First, to update the existing constraints by taking into account the latest mono-photon~\cite{Khachatryan:2014rwa, Aad:2014tda}, $\slashed{E}_T + W/Z \, (\to {\rm hadrons})$~\cite{Aad:2013oja} and VBF $h \to {\rm invisible}$~\cite{Chatrchyan:2014tja} searches. Second, to extend the studies~\cite{Cotta:2012nj,Carpenter:2012rg,Aad:2014vka,Nelson:2013pqa,Lopez:2014qja} by considering  in addition the $\slashed{E}_T  + W \, (\to {\rm leptons})$ channel~\cite{ATLAS:2014wra,Khachatryan:2014tva} as well as  the newest mono-jet data~\cite{Khachatryan:2014rra}. An assortment of Feynman diagrams that lead to the~$\slashed{E}_T$ signatures investigated in the following are displayed  in Fig.~\ref{fig:1}. 

Our analysis shows that depending on the choice of parameters, either the mono-photon or the mono-jet data give rise to the strongest restrictions at present. By combining the information on all available channels we are thus able to derive bounds on the coefficients $C_k (\Lambda)/\Lambda^3$ in (\ref{eq:1}) that improve on the existing limits. Building upon~\cite{Haisch:2013fla}, we furthermore demonstrate that measurements of the jet-jet azimuthal angle difference in $\slashed{E}_T + 2 j$ events may be used to disentangle whether the DM bilinear $\bar \chi \chi$ couples more strongly to the combination $B_{\mu \nu} B^{\mu \nu}$ ($W_{\mu \nu}^i W^{i, \mu \nu }$) or the product $B_{\mu \nu} \tilde B^{\mu \nu}$ ($W_{\mu \nu}^i \tilde W^{i, \mu \nu }$) of field strength tensors. Similar ideas have also been brought forward in \cite{Cotta:2012nj}.

The outline of this article is as follows. In Sec.~\ref{sec:2} we review the existing LHC searches for $\slashed{E}_T$ signatures that we will use to constrain the effective interactions~(\ref{eq:2}).  In~Sec.~\ref{sec:3} we derive the restrictions on the parameter space by combining all individual search modes, commenting also on how future measurements may improve these limits.  This section contains in addition a discussion of the azimuthal angle correlations between the two jets in the $\slashed{E}_T + 2j$ channel. Our conclusions are presented in Sec.~\ref{sec:4}.

\section{Search channels}
\label{sec:2}

In this section we list the various cuts and the values of the fiducial cross section ($\sigma_{\rm fid}$) of each individual~$\slashed{E}_T$ channel. This information will be used in the next section to set limits on the coefficients $C_k(\Lambda)/\Lambda^3$ appearing in the effective Lagrangian~(\ref{eq:1}).

\subsection{\boldmath Mono-photon signal}

We begin with the mono-photon signal, which has recently been searched for by both CMS~\cite{Khachatryan:2014rwa} and ATLAS~\cite{Aad:2014tda}. Since the former search leads to the stronger restrictions, we employ  the  CMS results, which are based on~$19.6 \, {\rm fb^{-1}}$ of $8 \, {\rm TeV}$ data. The relevant cuts are 
\begin{equation} \label{eq:3}
 \slashed{E}_T > 140 \, {\rm GeV} \,, \qquad |\eta_\gamma| <1.4442\,,
\end{equation}
where $\eta_\gamma$ denotes the pseudorapidity of the photon. The CMS collaboration performs the measurement in six different signal regions with a varying cut on the transverse momentum of the photon ($p_{T,\gamma}$). Note that due to the higher-dimensional nature of the operators~(\ref{eq:2}), the~$\slashed{E}_T + \gamma$ signal has a rather hard $p_{T,\gamma}$ spectrum. As a result, we find  that the most severe cut of $p_{T,\gamma} >  700 \, {\rm GeV}$ gives the strongest bounds on the parameter space in our case. The corresponding 95\% confidence level (CL) limit on the fiducial cross section reads 
\begin{equation}  \label{eq:4}
\sigma_{\rm fid} ( pp \to \slashed{E}_T + \gamma) < 0.22 \, {\rm fb} \,.
\end{equation}

\subsection{\boldmath Mono-$Z$ signal}

In the case of the $\slashed{E}_T + Z \, (\to \ell^+ \ell^-)$ search channel, we use the ATLAS results \cite{Aad:2014vka}, that utilise $20.3 \, {\rm fb}^{-1}$ of $8 \, {\rm TeV}$ data.  The selection criteria relevant to our analysis are
\begin{eqnarray} \label{eq:5}
\begin{split}
& p_{T,\ell} > 20 \, {\rm GeV} \,, \quad |\eta_\ell|<2.5 \,,  \quad   m_{\ell \ell} \in [76, 106] \, {\rm GeV} \,, \\[2mm]
&\phantom{xxxxxxx} |\eta_{\ell \ell}|<2.5 \,, \quad  \frac{| p_{T,\ell \ell} - \slashed{E}_T|}{ p_{T,\ell \ell}} < 0.5 \,.
\end{split}
\end{eqnarray}
Here  $m_{\ell \ell}$, $\eta_{\ell \ell}$ and $p_{T,\ell \ell}$ denote the invariant mass, the pseudorapidity and the transverse momentum of the di-lepton system, respectively. The ATLAS analysis defines four signal regions with different lower~$\slashed{E}_T$ thresholds. As it turns out, in the considered case the requirement~$\slashed{E}_T > 350 \, {\rm GeV}$ gives rise to the best bounds. Including $Z$-boson decays to both electrons and muons~($\ell = e,\mu$), the ATLAS experiment obtains for this~$\slashed{E}_T$ cut the following 95\%~CL bound
\begin{equation}  \label{eq:6}
\sigma_{\rm fid} \big ( pp \to \slashed{E}_T + Z \, (\to \ell^+ \ell^-) \big ) < 0.27 \, {\rm fb} \,.
\end{equation}

\subsection{\boldmath Mono-$W$ signal}

Both ATLAS~\cite{ATLAS:2014wra} and CMS~\cite{Khachatryan:2014tva} have searched for a mono-$W$  signal in the leptonic decay mode. We find that the ATLAS search for the $\mu \nu_\mu$ final state, which uses $20.3 \, {\rm fb}^{-1}$  of $8 \, {\rm TeV}$ data, gives the strongest constraints, and thus we consider only this channel. The most important experimental cuts are 
\begin{equation} \label{eq:7}
\begin{split}
&  p_{T,\mu} > 45 \, {\rm GeV} \,, \qquad |\eta_\mu| \in [0,1] \cup  [1.3,2] \,,  \\[2mm] 
& \phantom{xx} m_T = \sqrt{2 \hspace{0.25mm} p_{T,\mu} \slashed{E}_T \left ( 1 - \cos \varphi_{\mu  \slashed{E}_T} \right ) } \,,
\end{split}
\end{equation}
where $m_T$ is the transverse mass which depends on  the angle $\varphi_{\mu \slashed{E}_T} $ between the $p_{T,\mu}$ and the $\slashed{E}_T$ vectors. ATLAS sets bounds on $\sigma_{\rm fid}$ for three different $m_T$ cuts, and like in the case of the mono-photon signal, we observe that the strongest restriction of $m_T > 843 \, {\rm GeV}$ provides the best limits on the interactions~(\ref{eq:2}). At~95\% CL the  bound on the corresponding fiducial signal cross section is given by 
\begin{equation}  \label{eq:8}
\sigma_{\rm fid} \big ( pp \to \slashed{E}_T + W \, (\to \mu \nu_\mu) \big ) < 0.54 \, {\rm fb} \,.
\end{equation}

\subsection{\boldmath  Mono-$W/Z$ signal}

The ATLAS search \cite{Aad:2013oja} looks for a $\slashed{E}_T + W/Z$ signal, where the $W$ or $Z$ boson  decays hadronically.  This analysis is based on $20.3 \, {\rm fb}^{-1}$ of $8 \, {\rm TeV}$ data, jet candidates are reconstructed using the Cambridge/Aachen~(C/A) algorithm~\cite{Dokshitzer:1997in} with a radius parameter $R = 1.2$ and subjected to a mass-drop filtering procedure~\cite{Butterworth:2008iy}. Events are required to have at least one C/A jet with 
\begin{equation} \label{eq:9}
\begin{split}
& \phantom{ii} p_{T,j} > 250 \, {\rm GeV} \,, \qquad |\eta_j| < 1.2 \,, \\[2mm] 
& m_j \in [50, 120 ] \, {\rm GeV} \,, \qquad \sqrt{y} > 0.4 \,.
\end{split}
\end{equation}
Here $m_j$ refers to the mass of the large-radius jet, while $\sqrt{y} ={\rm min} \left (p_{T, j_1}, p_{T, j_2} \right ) \sqrt{(\Delta \phi_{j_1j_2})^2 + ( \Delta \eta_{j_1j_2})^2}/m_j$ is a measure of the momentum balance of the two leading subjets $j_1$ and $j_2$ contained in the C/A jet. The~95\%~CL limits on the fiducial cross section depend also on the imposed $\slashed{E}_T$ threshold, and it turns out that the stronger of the two cuts, i.e.~$\slashed{E}_T > 500 \, {\rm GeV}$, provides the most stringent constraints. In this case, the relevant limit on the fiducial cross section is 
\begin{equation}  \label{eq:10}
 \sigma_{\rm fid} \big ( pp \to \slashed{E}_T + W/Z \, (\to {\rm hadrons}) \big ) < 2.2 \, {\rm fb} \,.
\end{equation}

\subsection{\boldmath  Mono-jet  signal}

One can also use mono-jet events to constrain the operators in (\ref{eq:2}), since the corresponding searches allow for the presence of a secondary jet. Here we will  employ the newest CMS results \cite{Khachatryan:2014rra}, which make use of $19.7 \, {\rm  fb}^{-1}$ of $8 \, {\rm TeV}$ data. Like CMS, we reconstruct jets using an anti-$k_t$ algorithm~\cite{Cacciari:2008gp} with radius parameter $R = 0.5$. The relevant selection cuts are 
\begin{equation} \label{eq:11}
\begin{split}
p_{T,j_1} & > 110 \, {\rm GeV} \,,  \qquad |\eta_{j_1}| < 2.4 \,, \\[1mm]
p_{T,j_2} & > 30 \, {\rm GeV} \,, \qquad \; \, |\eta_{j_2}| < 4.5  \,, \\[2mm]
& \phantom{xxxi} \; \Delta \phi _{j_1j_2} < 2.5 \,,
\end{split}
\end{equation}
where $ \Delta \phi_{j_1j_2}$ is the azimuthal separation of the two leading jets. Another important selection criterion is the imposed jet-veto \cite{Haisch:2013ata}, which rejects  events  if they contain a tertiary  jet with $p_{T,j_3} > 30 \, {\rm GeV}$ and $|\eta_{j_3}| < 4.5$. The~CMS measurement is performed for seven different $\slashed{E}_T$ regions, and we find  that for the considered interactions the highest sensitivity is obtained for $\slashed{E}_T > 500 \, {\rm GeV}$. The corresponding~95\%~CL  limit on the fiducial cross section reads
\begin{equation}  \label{eq:12}
\sigma_{\rm fid} (pp \to \slashed{E}_T + 2 j ) < 6.1 \, {\rm fb} \,.
\end{equation}

\subsection{\boldmath  VBF invisible Higgs-boson decays}

Last but not least, we consider the results of the CMS search for invisible decays of the Higgs boson in the VBF channel \cite{Chatrchyan:2014tja}, which uses a $8 \, {\rm TeV}$ data sample, corresponding to an integrated luminosity of $19.5 \, {\rm fb}^{-1}$. Jets are reconstructed  employing an anti-$k_t$ clustering algorithm with $R=0.5$, and subject to the following requirements
\begin{equation}  \label{eq:13}
\begin{split}
&  p_{T,j_1}, p_{T,j_2} > 50 \, {\rm GeV} \,, \qquad |\eta_{j_1}|,  |\eta_{j_2}|  < 4.7 \,, \\[2mm]
& \phantom{xxxxi} \eta_{j_1} \cdot \eta_{j_2} < 0 \,, \qquad \Delta \eta_{j_1j_2} > 4.2 \,, \\[2mm] 
& \phantom{xxi} m_{j_1 j_2} > 1100 \, {\rm GeV} \,,  \qquad  \Delta \phi_{j_1 j_2} < 1.0 \,.
\end{split}
\end{equation}
The missing-energy cut is $\slashed{E}_T > 130 \, {\rm GeV}$ and a central jet-veto is imposed to any event that has a third  jet with $p_{T,j_3} > 30 \, {\rm GeV}$ and a pseudorapidity between those of the two tagging jets. For these cuts, CMS obtains the following 95\% CL bound on the fiducial cross section 
\begin{equation}  \label{eq:14}
\sigma_{\rm fid} (pp \to \slashed{E}_T + 2 j ) < 6.5 \, {\rm fb} \,.
\end{equation}

\section{Numerical results}
\label{sec:3}

\begin{figure}[!t]
\begin{center}
\includegraphics[width=0.45 \textwidth]{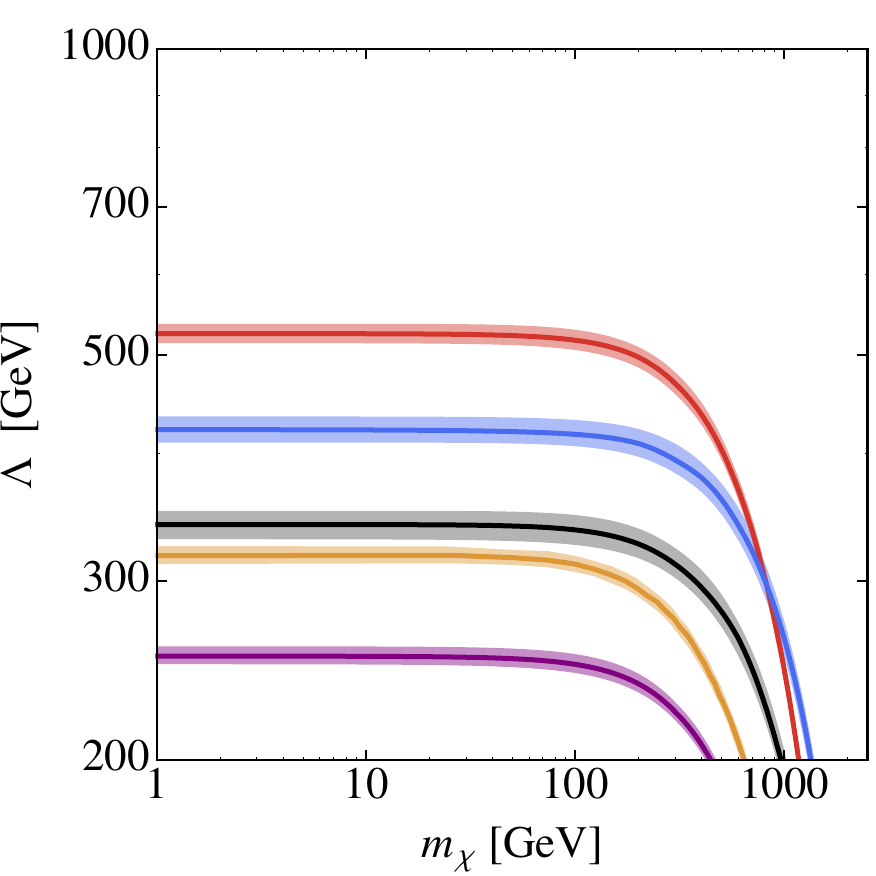} 
\includegraphics[width=0.45 \textwidth]{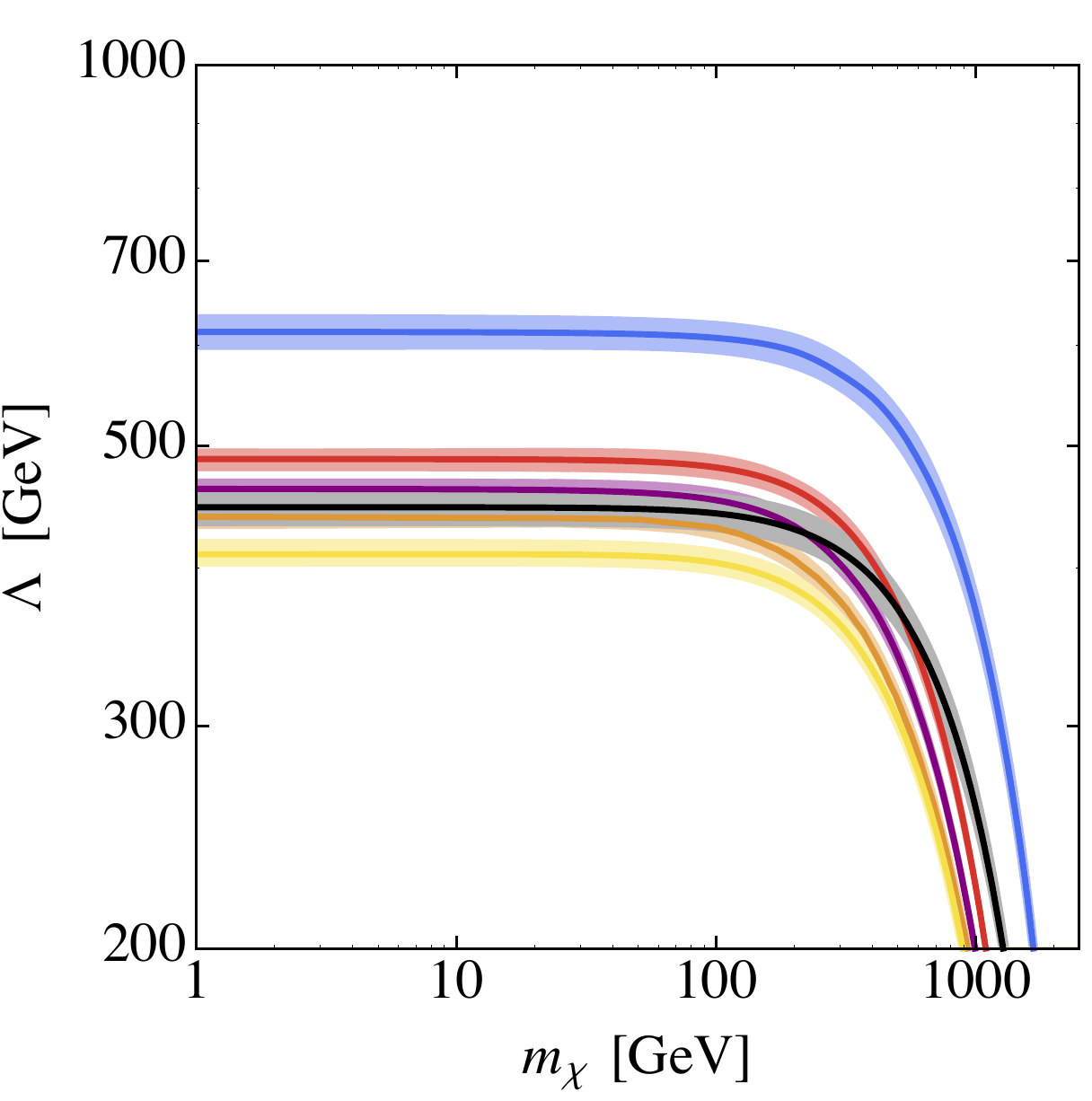} 
\caption{\label{fig:2} Assortment of LHC bounds on the new-physics scale~$\Lambda$, assuming $C_B (\Lambda)=1$, $C_W (\Lambda)=0$~(upper panel) and $C_B (\Lambda)=0$, $C_W (\Lambda)=1$~(lower  panel).  In both cases the DM particles are taken to be Dirac and $C_{\tilde B} (\Lambda)= C_{\tilde W} (\Lambda)=0$. The coloured  curves correspond to the limits arising from the latest mono-photon~(red), $\slashed{E}_T + Z \, (\to \ell^+ \ell^-)$~(orange), $\slashed{E}_T + W \, (\to \mu \nu_\mu)$~(yellow), $\slashed{E}_T + W/Z \, (\to {\rm hadrons})$~(purple),  mono-jet~(blue) and VBF~$h \to {\rm invisible}$~(grey) searches. The width of the bands reflect the associated scale uncertainties.}
\end{center}
\end{figure}

\begin{figure*}[!t]
\begin{center}
\includegraphics[width=0.32 \textwidth]{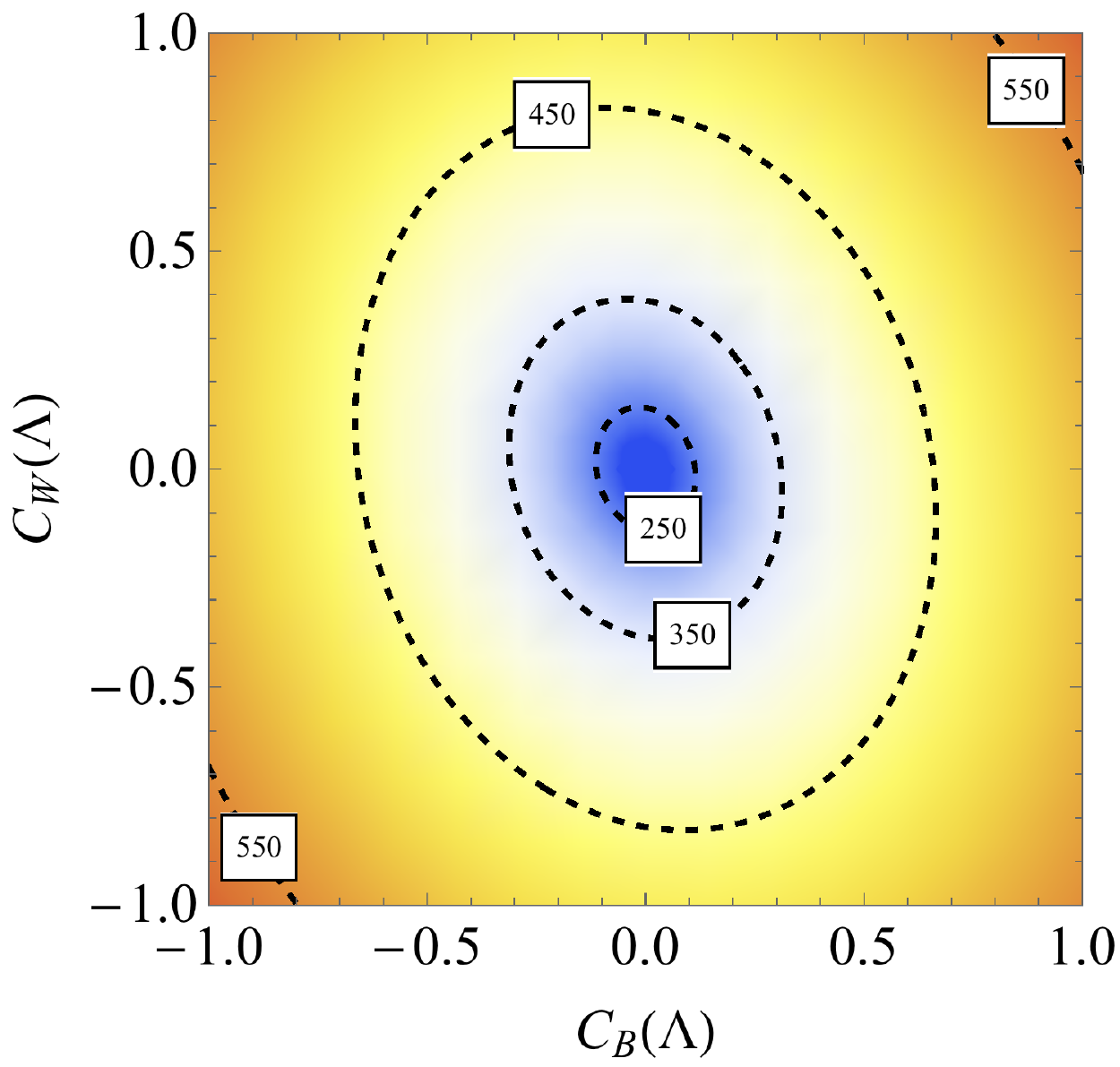} \, \includegraphics[width=0.32 \textwidth]{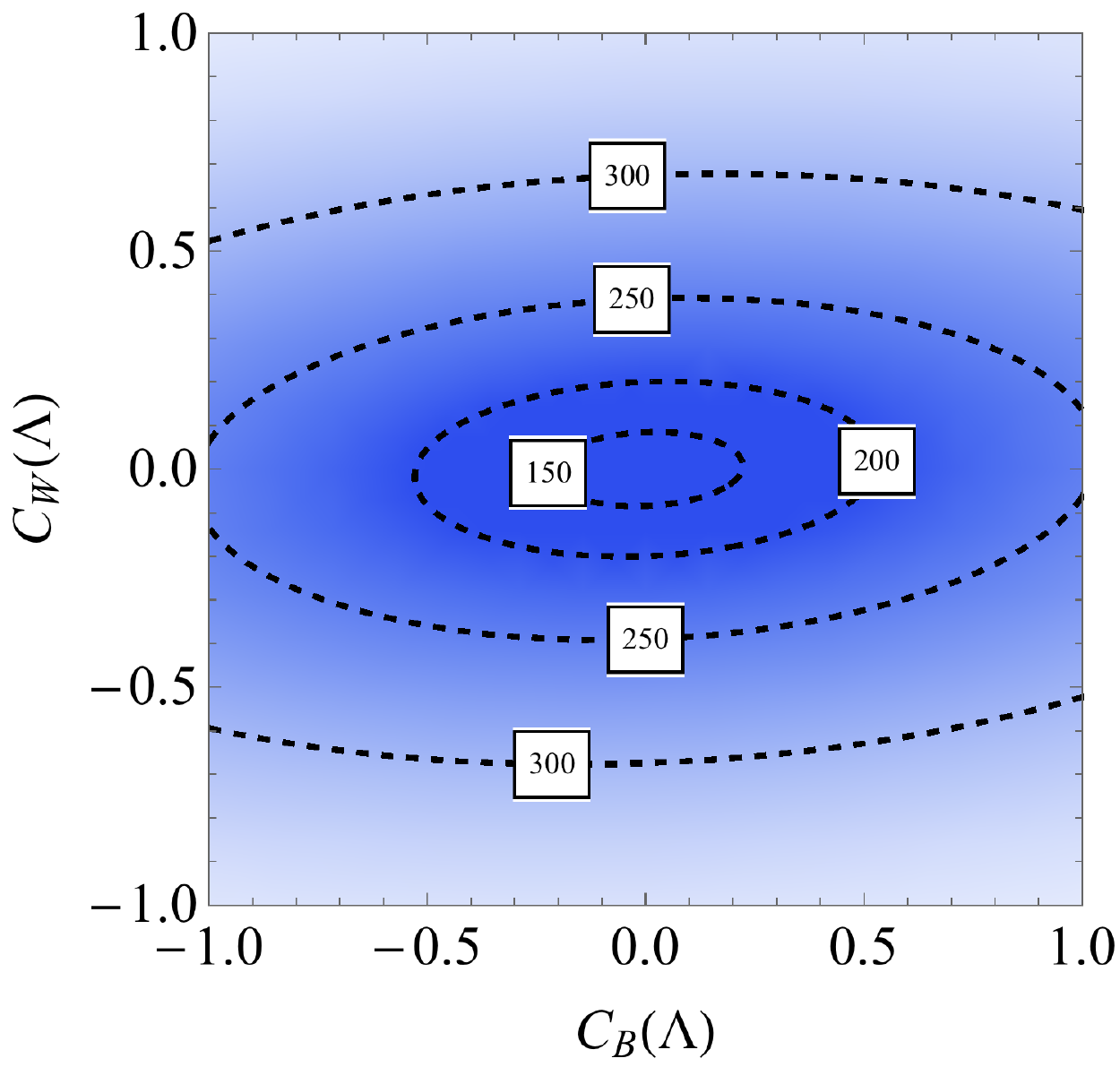}  \, \includegraphics[width=0.32 \textwidth]{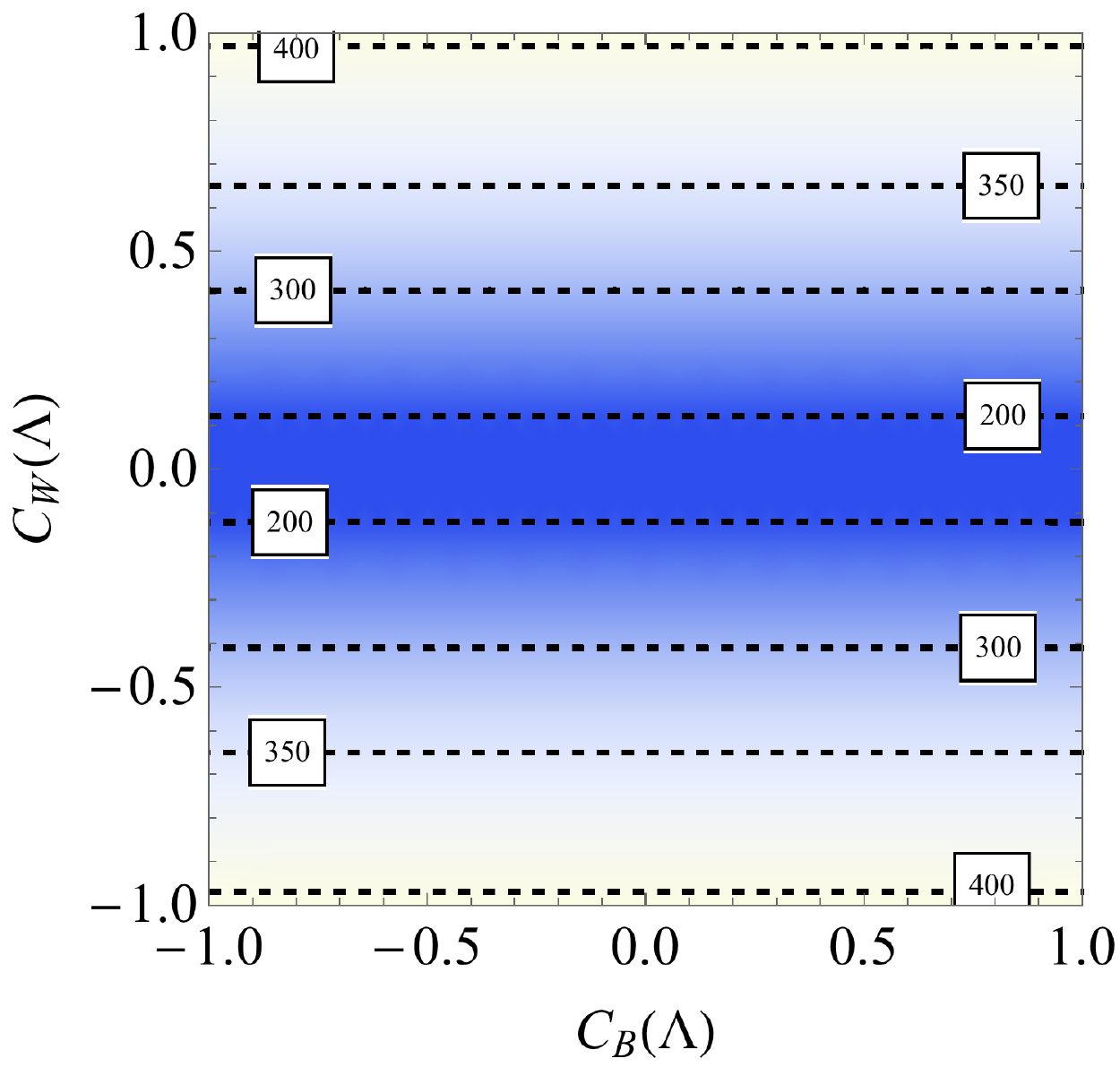} 

\vspace{2mm}

\includegraphics[width=0.32 \textwidth]{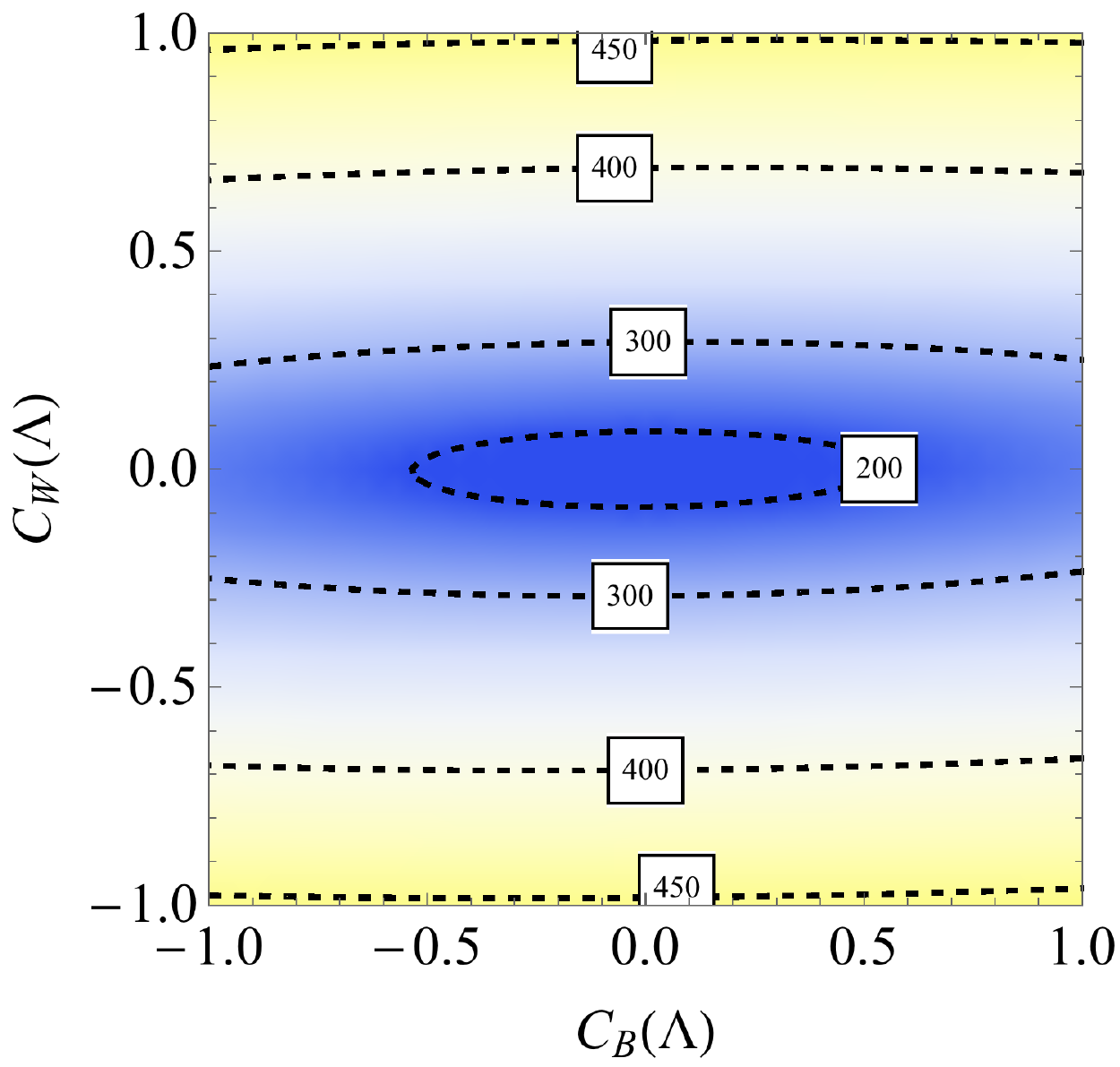}  \, \includegraphics[width=0.32 \textwidth]{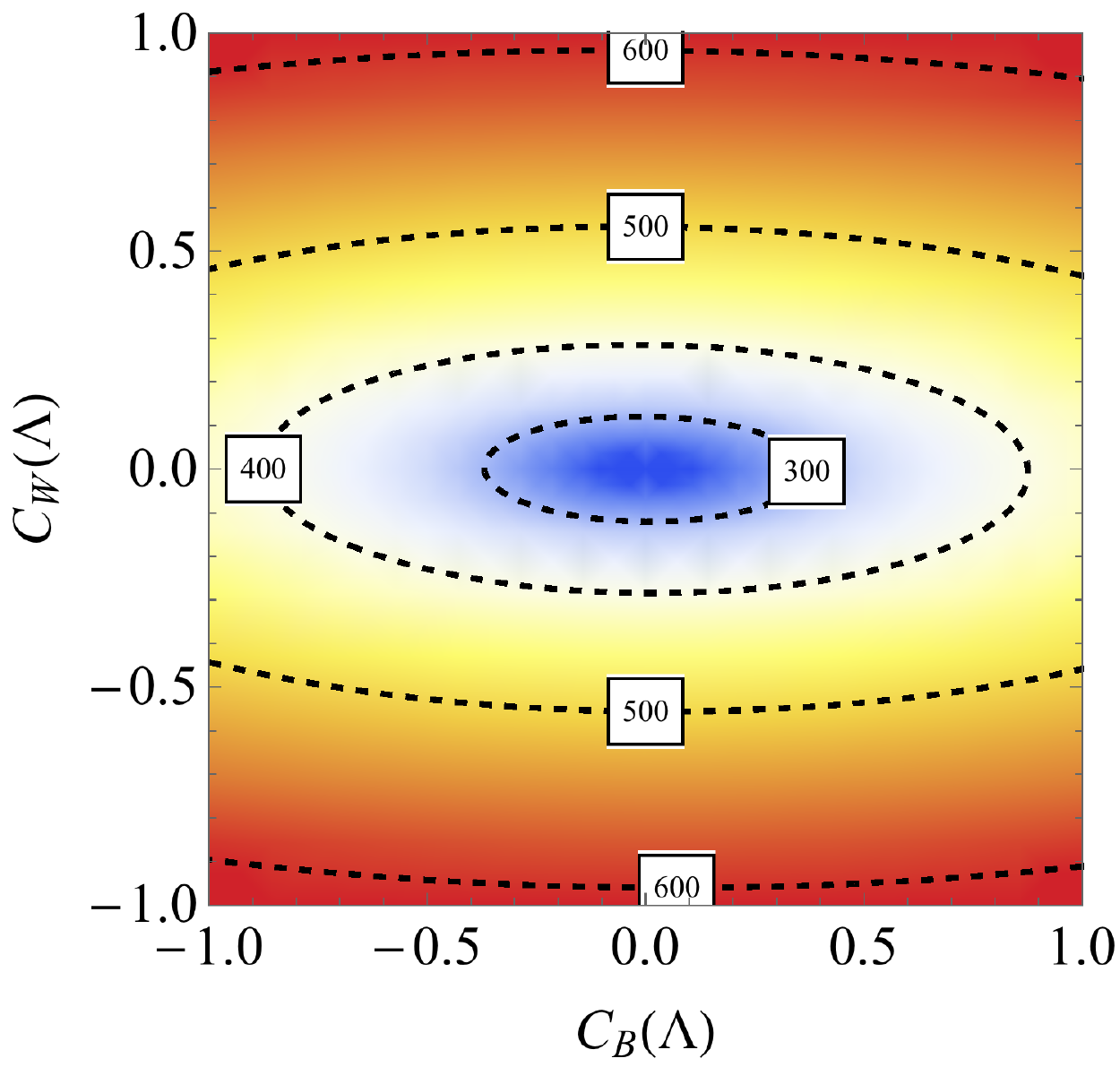} \, \includegraphics[width=0.325 \textwidth]{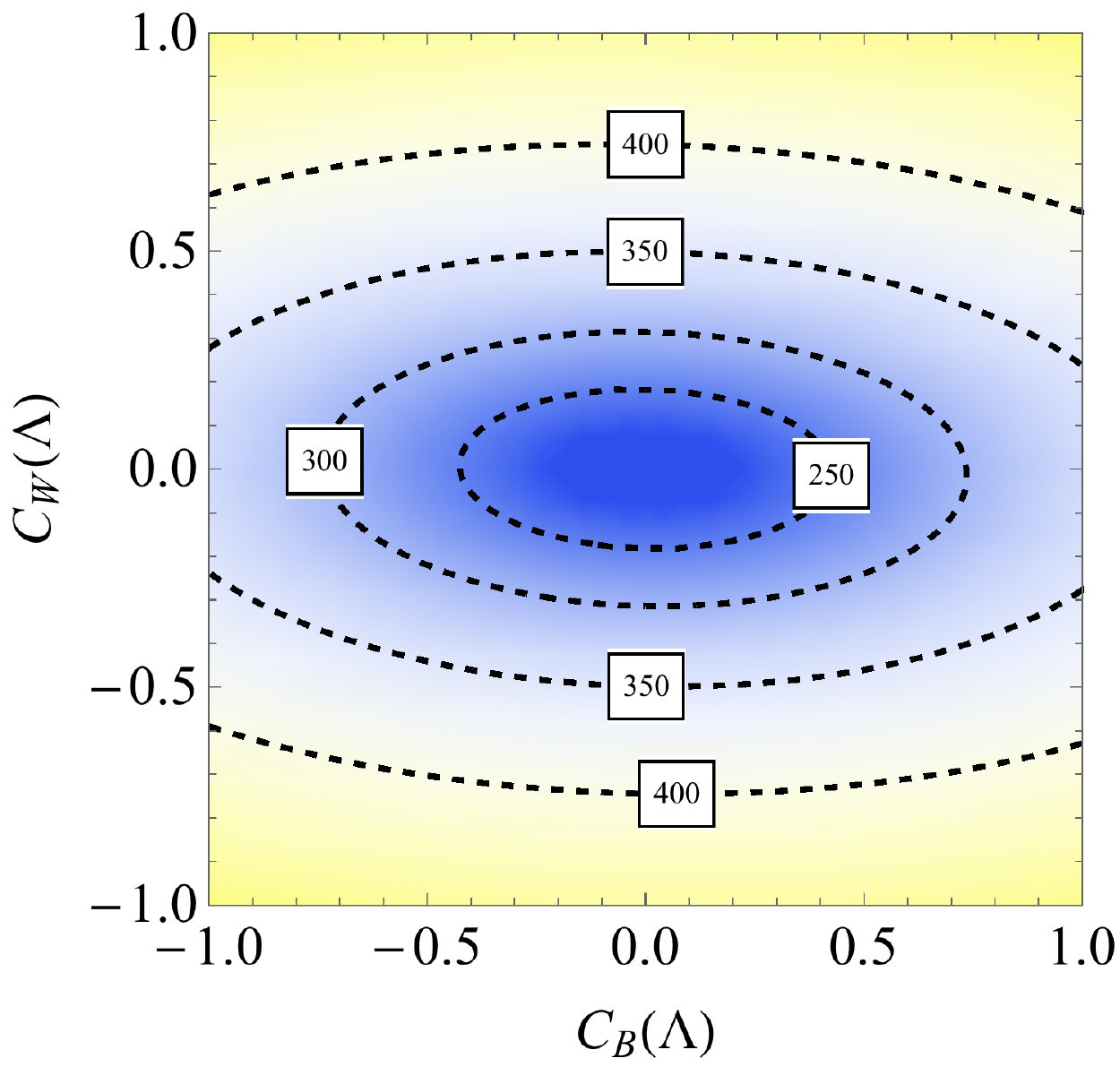} 
\caption{\label{fig:3} Limits on $\Lambda$ in the $C_B (\Lambda)\hspace{0.25mm}$--$\hspace{0.25mm}C_W (\Lambda)$ plane. The different panels correspond to the mono-photon~(upper left), $\slashed{E}_T + Z \, (\to \ell^+ \ell^-)$~(upper middle), $\slashed{E}_T + W \, (\to \mu \nu_\mu)$~(upper right),  $\slashed{E}_T + W/Z \, (\to {\rm hadrons})$~(lower left), mono-jet~(lower middle) and VBF $h \to {\rm invisible}$~(lower right) search. All results employ $m_\chi = 100 \, {\rm GeV}$ and $C_{\tilde B} (\Lambda) = C_{\tilde W} (\Lambda) =0$. The contour labels indicate the value of the  new-physics scale in units of ${\rm GeV}$.}
\end{center}
\end{figure*}

In order to determine the cross section for the $\slashed{E}_T$ signals associated to the effective operators~(\ref{eq:2}), we have implemented each of them in FeynRules~\cite{Christensen:2008py}, generating a~UFO output \cite{Degrande:2011ua}. The actual event generation has been performed at leading order with  MadGraph~5~\cite{Alwall:2011uj} utilising CTEQ6L1 parton distributions~\cite{Pumplin:2002vw}. Parton-shower effects and hadronisation corrections have been included by means of PYTHIA~8~\cite{Sjostrand:2007gs} and jets constructed using~FastJet~3~\cite{Cacciari:2011ma}. We employ Delphes~3~\cite{deFavereau:2013fsa} as a fast detector simulation to estimate the reconstruction efficiencies for the different $\slashed{E}_T$ signals. The efficiencies that we find amount to around 70\% for the mono-photon signal, 60\% in both the mono-$Z$ and mono-$W$ case and  65\% for the mono-$W/Z$ signature. These findings agree with~\cite{Aad:2014tda} for the $\slashed{E}_T + \gamma$,~\cite{Aad:2014vka} for the $\slashed{E}_T + Z$,~\cite{Lopez:2014qja} for the $\slashed{E}_T + W$ and~\cite{Aad:2013oja} for the $\slashed{E}_T +W/Z$ signal. For the mono-jet signal and the search for invisible decays of the Higgs boson in the VBF channel, we find reconstruction efficiencies in the ballpark of 95\%.

Our Monte Carlo (MC) implementation has been validated by reproducing the numerical results of~\cite{Aad:2014vka,Nelson:2013pqa} within theoretical uncertainties. These errors have been assessed by studying the scale ambiguities of our results. We have used the default dynamical scale choice of MadGraph~5, varying the scale factor in the  range $[1/2,2]$. We find that the predictions for the mono-photon, $\slashed{E}_T + Z \, (\to \ell^+ \ell^-)$ and  $\slashed{E}_T + W \, (\to \mu \nu_\mu)$  cross sections  calculated in this way vary by around~$\pm 15\%$, while in the case of the $\slashed{E}_T + W/Z\, (\to {\rm hadrons})$, the mono-jet and the VBF $h \to {\rm invisible}$ signal, relative differences of about~$\pm 20\%$ are obtained. Note that these errors are smaller than those found in~\cite{Haisch:2013ata,Haisch:2012kf,Haisch:2013uaa,Haisch:2015ioa}, since all the tree-level~$\slashed{E}_T$ cross sections considered in our work do not explicitly depend on $\alpha_s$. The quoted uncertainties thus reflect only the ambiguities related to the change of factorisation scale, but not renormalisation scale. 

\subsection{\boldmath  Dependence on a single Wilson coefficient}  

In Fig.~\ref{fig:2} we present the limits on the new-physics scale~$\Lambda$ for $C_{\tilde B} (\Lambda)= C_{\tilde W} (\Lambda)=0$ and the two choices  $C_B (\Lambda)=1$, $C_W (\Lambda)=0$~(upper panel) and $C_B (\Lambda)=0$, $C_W (\Lambda)=1$~(lower  panel) for the Wilson coefficients evaluated at $\Lambda$. The shown predictions correspond to Dirac DM and  the widths of the coloured bands illustrate the impact of scale variations.  For $C_B (\Lambda)=1$, $C_W (\Lambda)=0$, one observes that the mono-photon search~\cite{Khachatryan:2014rwa} provides the strongest constraints in most of the parameter space. Numerically, we find that the scale $\Lambda$ has to satisfy $\Lambda \gtrsim 510 \, {\rm GeV}$ for $m_\chi \lesssim 100 \, {\rm GeV}$ in order to meet the 95\%~CL requirement~(\ref{eq:4}). In the case~$C_B (\Lambda)=0$, $C_W (\Lambda)=1$, on the other hand, the latest mono-jet data~\cite{Khachatryan:2014rra} impose the leading restrictions.  At 95\%~CL, the inequality  (\ref{eq:12}) translates into a lower limit of $\Lambda \gtrsim 600 \, {\rm GeV}$ for DM masses below $100 \, {\rm GeV}$. The shown  limits also hold in the  case that $C_{\tilde B} (\Lambda)=1$, $C_{\tilde W} (\Lambda)=0$ or $C_{\tilde B} (\Lambda)=0$, $C_{\tilde W} (\Lambda)=1$ and $C_B (\Lambda)=C_W (\Lambda)=0$, while for Majorana DM the constraints on $\Lambda$ would be stronger by around $12\%$. Note finally that $\slashed{E}_T + W \, (\to \mu \nu_\mu)$ searches do not provide any constraint on scenarios with $C_W (\Lambda) = C_{\tilde W} (\Lambda)=0$.  

To better understand the restrictions imposed by the various search channels, we consider the Feynman rules associated to the effective operators $O_B$ and $O_W$ entering (\ref{eq:1}). In momentum space, the resulting interactions between pairs of DM particles and SM gauge bosons  take the form 
\begin{equation}  \label{eq:15}
\frac{4 \hspace{0.25mm} i}{\Lambda^3} \; g_{V_1 V_2} \, \big (  p_1^{\mu_2} \hspace{0.25mm} p_2^{\mu_1} - g^{\mu_1 \mu_2}  \, p_1 \cdot p_2 \big ) \,,
\end{equation}
where $p_i$ ($\mu_i$) denotes the momentum (Lorentz index) of the vector field $V_i$ and for simplicity the spinors associated with the DM fields have been dropped. In terms of the sine ($s_w$) and cosine  ($c_w$) of the weak mixing angle and the Wilson coefficients $C_B(\Lambda)$ and $C_W(\Lambda)$, the couplings $g_{V_i V_j}$ read
\begin{equation} \label{eq:16}
\begin{split}
g_{AA} & = c_w^2 \hspace{0.25mm} C_B (\Lambda)+ s_w^2  \hspace{0.25mm} C_W (\Lambda)\,, \\[2mm]
& \hspace{-9mm} g_{AZ}   = - s_w c_w \, \big (  C_B (\Lambda) - C_W (\Lambda) \big ) \,, \\[2mm]
g_{ZZ}  & = s_w^2 \hspace{0.25mm} C_B (\Lambda)+ c_w^2  \hspace{0.25mm} C_W (\Lambda) \,, \\[2.5mm]
& \hspace{5mm} g_{WW} = C_W (\Lambda)\,.
\end{split}
\end{equation}
These results do not coincide with the expressions reported  in \cite{Carpenter:2012rg,Nelson:2013pqa,Lopez:2014qja}. From (\ref{eq:16}) we see that in the coupling  $g_{AA}$ of DM to two photons, the Wilson coefficients $C_B (\Lambda)$ enters compared to $C_W (\Lambda)$ with a relative factor of $c_w^2/s_w^2 \simeq 3.3$. On the other hand, in the case of the coupling between DM and $Z$-boson pairs $g_{ZZ}$, the dependence on $s_w$ and $c_w$ is reversed compared to $g_{AA}$. These properties explain why the limit on the new-physics scale $\Lambda$  from mono-photon $\big($$\slashed{E}_T + Z \, (\to \ell^+ \ell^-)$, $\slashed{E}_T + W/Z \, (\to {\rm hadrons})$, mono-jet and VBF $h \to {\rm invisible}$$\big)$ searches is  stronger (weaker) in the upper panel than in the lower panel of  Fig.~\ref{fig:2}. 

\begin{figure}[!t]
\begin{center}
\includegraphics[width=0.45 \textwidth]{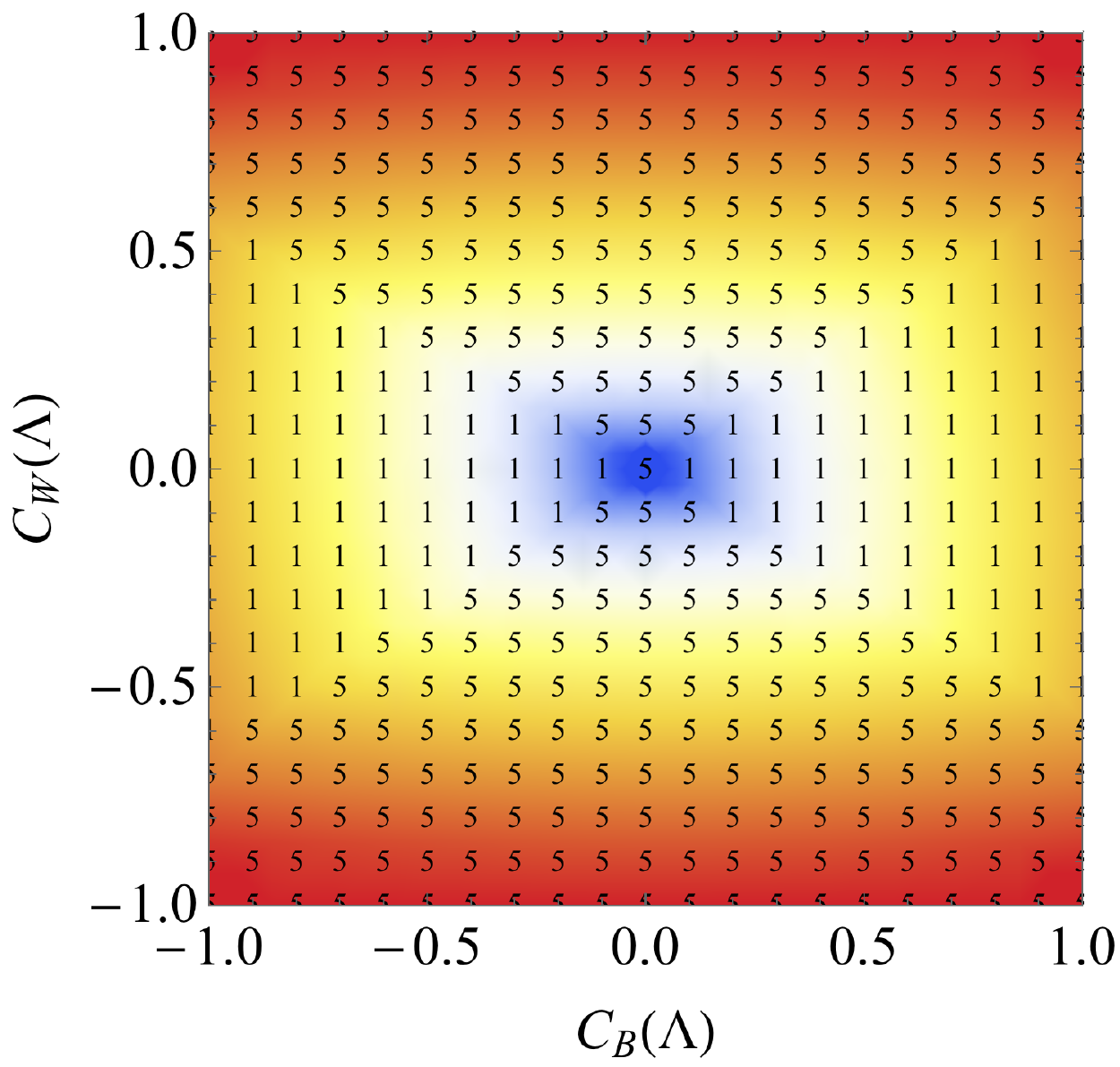} 

\vspace{2mm}

\includegraphics[width=0.45 \textwidth]{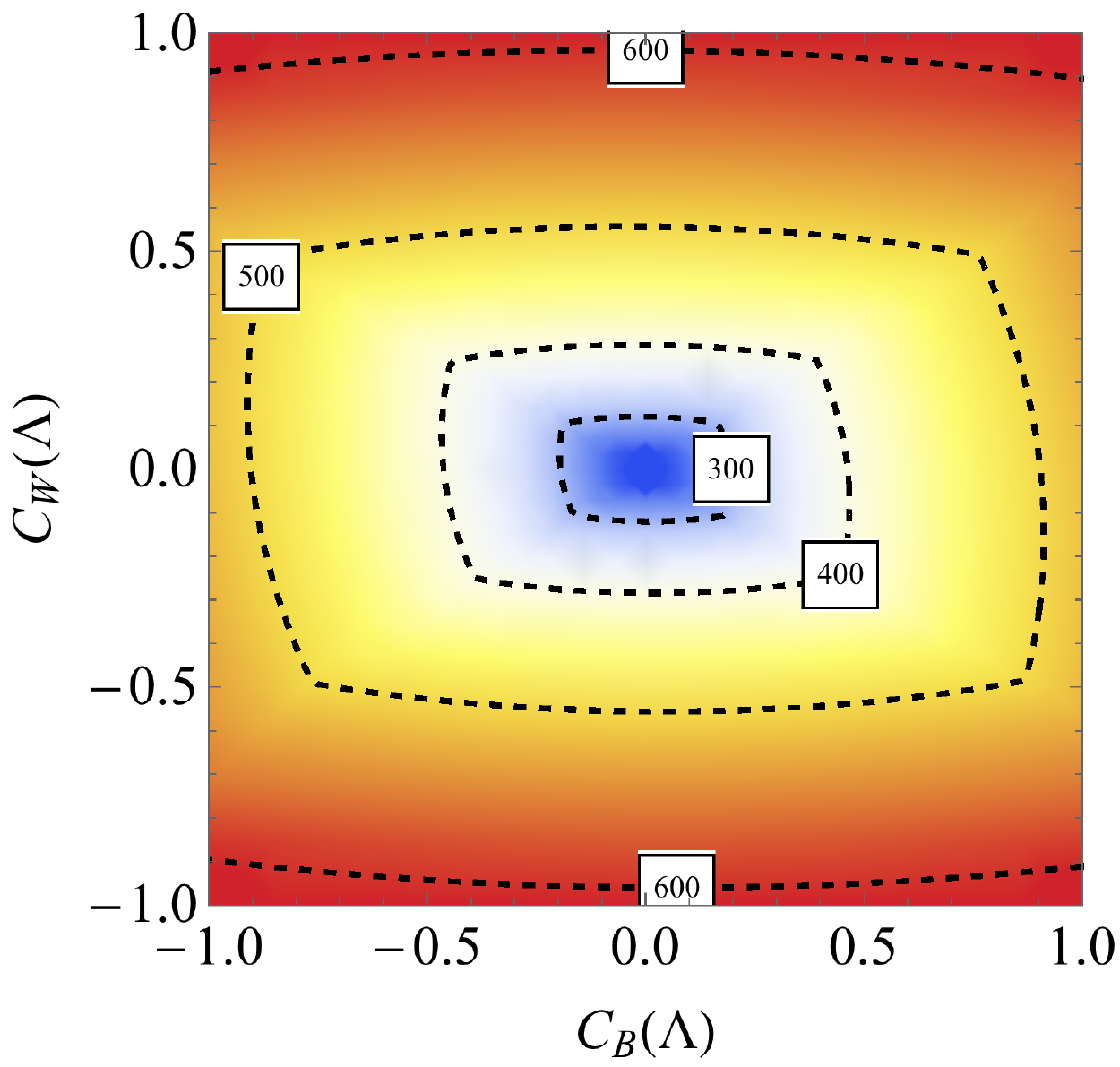} 
\caption{\label{fig:4}  Combination of the bounds on the new-physics scale in the $C_B (\Lambda)\hspace{0.25mm}$--$\hspace{0.25mm}C_W (\Lambda)$ plane, employing $m_\chi = 100 \, {\rm GeV}$ and $C_{\tilde B} (\Lambda) = C_{\tilde W} (\Lambda) =0$. In the upper panel  the search strategy that provides the leading constraint is indicated by the superimposed numbers, with 1 (5) representing the latest mono-photon (mono-jet) search, while the lower panel shows the resulting contours of constant  $\Lambda$   in units of $\rm GeV$.}
\end{center}
\end{figure}

A second important feature worth noting is that channels with leptons in the final state typically lead to weaker restrictions on the parameter space than modes involving hadrons. This is a simple consequence of the fact that the electroweak SM gauge bosons dominantly decay  hadronically. Numerically, one has ${\rm Br} \left ( Z \to \ell^+ \ell^- \right ) \simeq 7\%$ and ${\rm Br} \left ( W \to \mu \nu_\mu \right ) \simeq 11\%$, while ${\rm Br} \left ( Z \to {\rm hadrons} \right ) \simeq 70\%$ and ${\rm Br} \left ( W \to {\rm hadrons} \right ) \simeq 68\%$~\cite{Agashe:2014kda}.  The strong suppression of the leptonic decay widths  overcompensates the higher detection efficiencies of final states involving leptons, and as a result the LHC searches for $\slashed{E}_T + {\rm hadrons}$  are superior to those looking for $\slashed{E}_T + {\rm leptons}$ signals. 

Our third observation is that the latest mono-jet data are evidently more constraining than the recent VBF $h \to {\rm invisible}$ search. While these analyses explore the same final state, i.e.~$\slashed{E}_T + 2 j$, they probe quite different parts of the phase space. In fact, the selection criterion that has the biggest impact in our study is the rather loose missing transverse energy cut of  $\slashed{E}_T > 130 \, {\rm GeV}$ imposed in the VBF $h \to {\rm invisible}$ search. This selection is tailored for a Higgs boson of $125 \, {\rm GeV}$, but fares less well if one tries to probe higher-dimensional operators of the form (\ref{eq:2}). Since the operators $O_B$ ($O_{\tilde B}$) and $O_W$ ($O_{\tilde W}$) produce a rather hard $\slashed{E}_T$  spectrum, more severe  $\slashed{E}_T$ requirements allow for a cleaner separation between signal and SM background. 

\subsection{\boldmath  Dependence on two Wilson coefficients}  

Until now we have have studied the constraints on the new-physics scale $\Lambda$ as a function of the DM mass $m_\chi$, keeping the values of the high-scale Wilson coefficients fixed. In the panels of Fig.~\ref{fig:3} we instead show contours of constant~$\Lambda$ in the $C_B (\Lambda)\hspace{0.25mm}$--$\hspace{0.25mm}C_W (\Lambda)$ plane. In all plots we employ $m_\chi = 100 \, {\rm GeV}$ and set $C_{\tilde B} (\Lambda) = C_{\tilde W} (\Lambda) =0$.  The first noticeable feature of the shown predictions is that only the mono-photon signal depends more strongly on $C_B (\Lambda)$ than $C_W (\Lambda)$, while for all the other $\slashed{E}_T$ channels the situation is reversed. Second, with the exception of the mono-photon case, one observes that the major axes of the elliptic contours in all panels are almost aligned with the $C_W(\Lambda)$ axes. This means that interference effects between contributions arising from $O_B$ and $O_W$ are small in all of these cases. The third important property following from the colour shading of the depicted results is that currently either the newest mono-photon or the mono-jet data provide the leading bounds in the entire $C_B (\Lambda)\hspace{0.25mm}$--$\hspace{0.25mm}C_W (\Lambda)$ plane. This feature is further illustrated by the upper panel in Fig.~\ref{fig:4}. In this plot the overlaid numbers indicate the search strategy that contributes the best sensitivity on $\Lambda$ at each point, with 1 and 5 corresponding to the mono-photon and mono-jet  channel, respectively. One sees that if the ratio of  Wilson coefficients satisfies $| C_B (\Lambda) /C_W (\Lambda)| \gtrsim 1.5$ then the limit~(\ref{eq:4}) gives rise to the strongest constraint, while in the remaining  $C_B (\Lambda)\hspace{0.25mm}$--$\hspace{0.25mm}C_W (\Lambda)$ plane the bound~(\ref{eq:14})  is most restrictive. The $\Lambda$ contours obtained by combining all available $\slashed{E}_T$ channels are depicted in the lower panel of Fig.~\ref{fig:4}. 

Finally, we note that the values of  $\Lambda$ that can be excluded with the current data are low compared to typical LHC energies. In order to go beyond the  EFT description, one has to specify a  ultraviolet (UV) completion, where the operators in (\ref{eq:4}) arise from a renormalisable theory after integrating out the heavy degrees of freedom mediating the interactions. UV-complete models that generate the operators $O_B$ and  $O_W$ through loops of states charged under $U(1)_Y$ and/or $SU(2)_L$  have been proposed in \cite{Weiner:2012gm} and their LHC signatures have been studied in \cite{Liu:2013gba}. If these new charged particles  are  light, the high-$p_T$ gauge bosons that participate in  the $\slashed{E}_T$ processes considered here are able to resolve the substructure of the loops. This generically suppresses the cross sections compared to the EFT predictions~\cite{Haisch:2012kf}, and thus will weaken the bounds on the interaction strengths of  DM and the electroweak gauge bosons  to some extent.  Furthermore, the light charged mediators may be produced  on-shell in $pp$ collisions, rendering direct LHC searches potentially more restrictive than $\slashed{E}_T$ searches. Making the above statements precise would require a study of a concrete UV completion.

\subsection{Future sensitivity}

It is also interesting to explore how the reach on the new-physics scale $\Lambda$ might improve at the  $14 \, {\rm TeV}$ LHC. In what follows, we will only consider the mono-jet signal, applying the event selection criteria that have been used in the sensitivity study by ATLAS~\cite{ATL-COM-PHYS-2014-549}. These read 
\begin{equation} \label{eq:17}
\begin{split}
p_{T,j_1} & > 300 \, {\rm GeV} \,,  \qquad |\eta_{j_1}| < 2.0 \,, \\[1mm]
p_{T,j_2} & > 50 \, {\rm GeV} \,, \qquad \; \, |\eta_{j_2}| < 3.6  \,, \\[2mm]
& \phantom{xxxi} \; \Delta \phi _{j \slashed{E}_T} > 0.5 \,,
\end{split}
\end{equation}
and jets are reconstructed using an anti-$k_t$ algorithm with $R = 0.4$. Events with a third jet of $p_{T,j_3}  > 50 \, {\rm GeV}$ and  $ |\eta_{j_3}| < 3.6$ are vetoed and the missing transverse energy cut that we employ is $\slashed{E}_T > 800 \, {\rm GeV}$. Note that compared to (\ref{eq:11}) the $p_{T, j_1}$, $p_{T, j_2}$ and $\slashed{E}_T$ thresholds are increased both to avoid pile-up and to enhance the signal-over-background ratio. In order to determine the limits on the scale $\Lambda$, we take $\sigma_{\rm fid} \big (pp \to Z \, (\to \bar \nu \nu) + j  \big )= 5.5 \, {\rm fb}$~\cite{ATL-COM-PHYS-2014-549}, assuming a total systematic uncertainty on the SM background of 5\%. For the choice $C_B (\Lambda) = 0$, $C_W (\Lambda) = 1$ and $C_{\tilde B} (\Lambda) = C_{\tilde W} (\Lambda) =0$, we find that with $25 \, {\rm fb^{-1}}$ of data, corresponding to the first year of running after the LHC upgrade to $14 \, {\rm TeV}$, one may be able  to set a 95\% CL bound of   $\Lambda \gtrsim 1.3 \, {\rm TeV}$ for $m_\chi \lesssim 100 \, {\rm GeV}$. Compared to the present limit, this corresponds to an improvement of the bound on $\Lambda$ by more than a factor of~2. With $300 \, {\rm fb}^{-1}$ and $3000  \, {\rm fb}^{-1}$ of accumulated data, we obtain instead $\Lambda \gtrsim 1.4 \, {\rm TeV}$. These numbers make clear that at $14 \, {\rm TeV}$ the sensitivity of $\slashed{E}_T + j$ searches will rather soon be limited by systematic uncertainties associated to the irreducible SM background. To what extent this limitation can be evaded by an optimisation of the mono-jet searches and/or an improved  understanding of the $pp \to Z \, (\to \bar \nu \nu) + j$ channel,  would require a dedicated study. Such an analysis is beyond the scope of this work. 

\subsection{Analysis of jet-jet angular correlations}

So far we have analysed only observables that are insensitive to whether the $\slashed{E}_T$ signal is generated by an insertion of the effective operator $O_B$ ($O_W$) or $O_{\tilde B}$ ($O_{\tilde W}$). This ambiguity can however be resolved by measuring the azimuthal angle difference $\Delta \phi_{j_1 j_2}$ of forward jets produced in $\slashed{E}_T + 2 j$ events~\cite{Cotta:2012nj,Haisch:2013fla}.  Besides the cuts~(\ref{eq:17}), we impose the following VBF-like selection requirements in our analysis 
\begin{equation}  \label{eq:18}
\begin{split}
& \eta_{j_1} \cdot \eta_{j_2} < 0 \,, \quad \Delta \eta_{j_1j_2} > 2\,,  \quad m_{j_1 j_2} > 1100 \, {\rm GeV}  \,.
\end{split}
\end{equation}
Here the cut on the pseudorapidity separation  helps to sculpt the angular correlations between the  tagging jets, while the di-jet invariant mass threshold improves the signal-over-background ratio. 

In order to understand why the operators $O_B$ ($O_W$) and $O_{\tilde B}$ ($O_{\tilde W}$) lead to different jet-jet angular correlations, one has to consider their Feynman rules. In the case of the operators containing regular field strength tensors this has already been done in (\ref{eq:15}), while for their dual counterparts we obtain 
\begin{equation}  \label{eq:19}
\frac{2 \hspace{0.25mm} i}{\Lambda^3} \; g_{V_1 V_2} \, \epsilon^{\mu_1 \mu_2 \nu  \hspace{0.25mm}  \lambda} \, \big (  p_{1  \hspace{0.25mm}  {\nu}} \hspace{0.25mm} p_{2  \hspace{0.25mm}  {\lambda}} -  p_{1  \hspace{0.25mm}  {\lambda}} \hspace{0.25mm} p_{2  \hspace{0.25mm}  {\nu}}    \big ) \,,
\end{equation}
with   $g_{V_i V_j}$ given in (\ref{eq:16}). The selection cuts (\ref{eq:18}) emphasise the parts of the phase space where the external partons experience only a small energy loss and the momentum components of the tagging jets in the beam direction are much greater than those in the transverse plane.  In this limit the structure of the $pp \to \slashed{E}_T + 2 j$ matrix elements  is straightforward to work out~\cite{Plehn:2001nj}. In the case of the effective operator $O_W$, one gets for instance ${\cal M}_W \sim   J_1^{\mu_1} J_2^{\mu_2} \hspace{0.5mm} (g_{\mu_1 \mu_2} \, p_1 \cdot p_2 - p_{1  \hspace{0.25mm}  \mu_1}  \hspace{0.25mm}  p_{2  \hspace{0.25mm}  \mu_2}) \sim \vec{p}_{T,j_1} \cdot \vec{p}_{T,j_2}$, while for~$O_{\tilde W}$ one arrives  instead at ${\cal M}_{\tilde W} \sim   \epsilon_{\mu_1 \mu_2  \nu \hspace{0.25mm} \lambda} \hspace{0.5mm} J_1^{\mu_1} J_2^{\mu_2} \hspace{0.25mm}  p_1^\nu \hspace{0.5mm} p_2^\lambda \sim \vec{p}_{T,j_1} \times \vec{p}_{T,j_2}$. Here  $J_{i}$ and $p_{i}$ denote the currents and momenta of the electroweak gauge bosons that partake in the scattering. These simple arguments imply that the $\Delta \phi_{j_1 j_2}$ spectrum corresponding to $O_W$ should be enhanced for collinear tagging jets, $\Delta \phi_{j_1 j_2} =0$, while for  $\Delta \phi_{j_1 j_2} =\pi/2$ it should show  an approximate zero. In the case of $O_{\tilde W}$, on the other hand, the  $\Delta \phi_{j_1 j_2} $ distribution should have a dip if the two jets are collinear, $\Delta \phi_{j_1 j_2} =0$, or back-to-back, $\Delta \phi_{j_1 j_2} = \pi$.  Note that the above arguments do not depend on the chirality of the DM current. This means that $O_B$, $O_W$, $O_{\tilde B}$, $O_{\tilde W}$ and the operators obtained from (\ref{eq:2}) by replacing $\bar \chi \chi$ with $\bar \chi \gamma_5 \chi$ lead to very similar jet-jet angular correlations, as we have explicitly verified.

\begin{figure}[!t]
\begin{center}
\includegraphics[width=0.45 \textwidth]{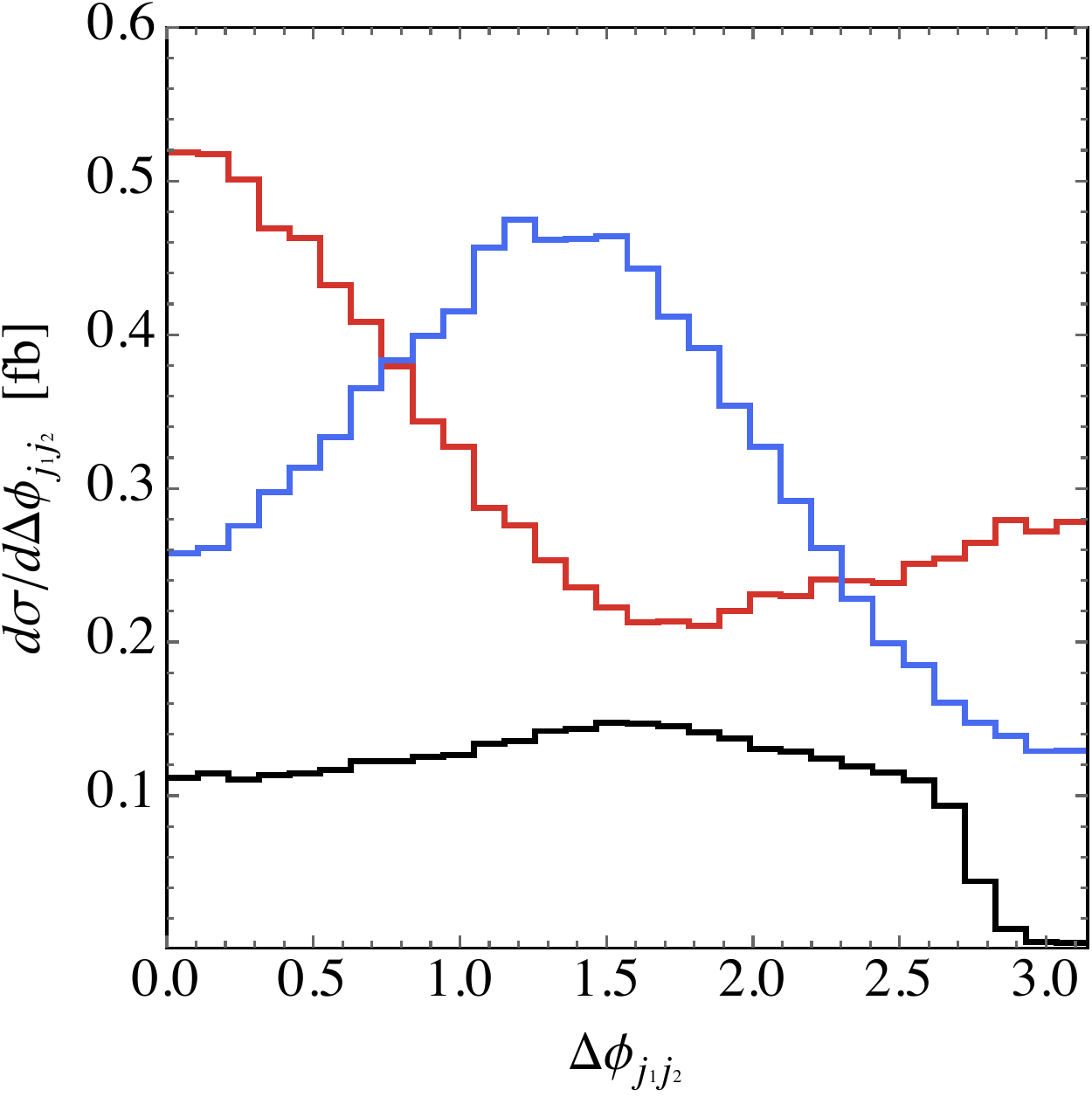} 
\caption{\label{fig:5}  Azimuthal angle distributions at the $14 \, {\rm TeV}$ LHC. The signal curves correspond to $C_B(\Lambda)=0$, $C_W(\Lambda)=1$~(red) and $C_{\tilde B}(\Lambda)=0$, $C_{\tilde W}(\Lambda)=1$ (blue), and both use  $\Lambda = 1 \, {\rm TeV}$ and $m_\chi = 100 \, {\rm GeV}$. For comparison the prediction of the dominant SM background process $pp \to Z \, (\to \bar \nu \nu) + 2 j$  (black) employing the same event selection criteria is shown as well.}
\end{center}
\end{figure}
 
In Fig.~\ref{fig:5} we plot the $\Delta \phi_{j_1 j_2}$ spectra for the choices $C_B(\Lambda)=0$, $C_W(\Lambda)=1$ (red curve) and $C_{\tilde B}(\Lambda)=0$, $C_{\tilde W}(\Lambda)=1$ (blue curve). All shown predictions are obtained for the $14 \, {\rm TeV}$ LHC and employ $\Lambda = 1 \, {\rm TeV}$ and  $m_\chi = 100 \, {\rm GeV}$. The  fiducial signal cross sections amount to $1.0 \, {\rm fb}$, independently of whether the insertion of~$O_W$ or $O_{\tilde W}$ is considered. The expected sine-like (cosine-like) behaviour of the modulation in the azimuthal angle distribution corresponding to $O_W$~($O_{\tilde W}$) is clearly visible  in the figure. These shapes should be contrasted with the spectrum of the dominant SM background process $pp \to Z \, (\to \bar \nu \nu) + 2 j$~(black curve),  which is rather flat for values $\Delta \phi_{j_1 j_2} \lesssim 2.6$ and then rapidly drops to zero. The corresponding fiducial cross section is  $0.35  \, {\rm fb}$, implying a signal-over-background ratio of $S/\sqrt{B} \simeq 8.4$, $29$ and $93$ for $25 \, {\rm fb}^{-1}$, $300 \, {\rm fb}^{-1}$  and $3000 \, {\rm fb}^{-1}$ of data, respectively.

The  given  $S/\sqrt{B} $ values imply that running the LHC for a couple of years at $14 \, {\rm TeV}$ should provide a sufficient number of events to analyse the  jet-jet  angular correlations. To quantify this statement, we use a  toy MC and generate event samples for both signals and background corresponding to $300 \, {\rm fb}^{-1}$ and $3000 \, {\rm fb}^{-1}$ of luminosity.  The resulting  differential cross sections are then fitted  to~\cite{Hankele:2006ja}
\begin{equation}  \label{eq:20}
\frac{1}{\sigma} \frac{d \sigma}{d \Delta \phi_{j_1 j_2}} = \sum_{n=0}^2 a_n  \hspace{0.25mm} \cos \left ( n \hspace{0.25mm} \Delta \phi_{j_1 j_2} \right ) \,.
\end{equation}
The coefficient $a_0 $ is fixed  by the normalisation of the $ \Delta \phi_{j_1 j_2}$ spectrum, and the ratio $r_1 = a_1/a_0$ turns out to be rather insensitive to which type of higher-dimensional interactions is considered. In contrast, the combination $r_2 = a_2/a_0$ is a measure of the CP nature of the interactions that lead to the $2j$ final state (see~e.g.~\cite{Plehn:2001nj,Hankele:2006ja}). This ratio is expected to be positive (negative) for an insertion of $O_B$ ($O_{\tilde B}$) and $O_W$ ($O_{\tilde W}$). We  stress that by considering normalised $\Delta \phi_{j_1 j_2}$ distributions, theoretical uncertainties are reduced and that the predictions become fairly  independent of  EFT assumptions~\cite{Haisch:2013fla}.

\begin{figure}[!t]
\begin{center}
\includegraphics[width=0.45 \textwidth]{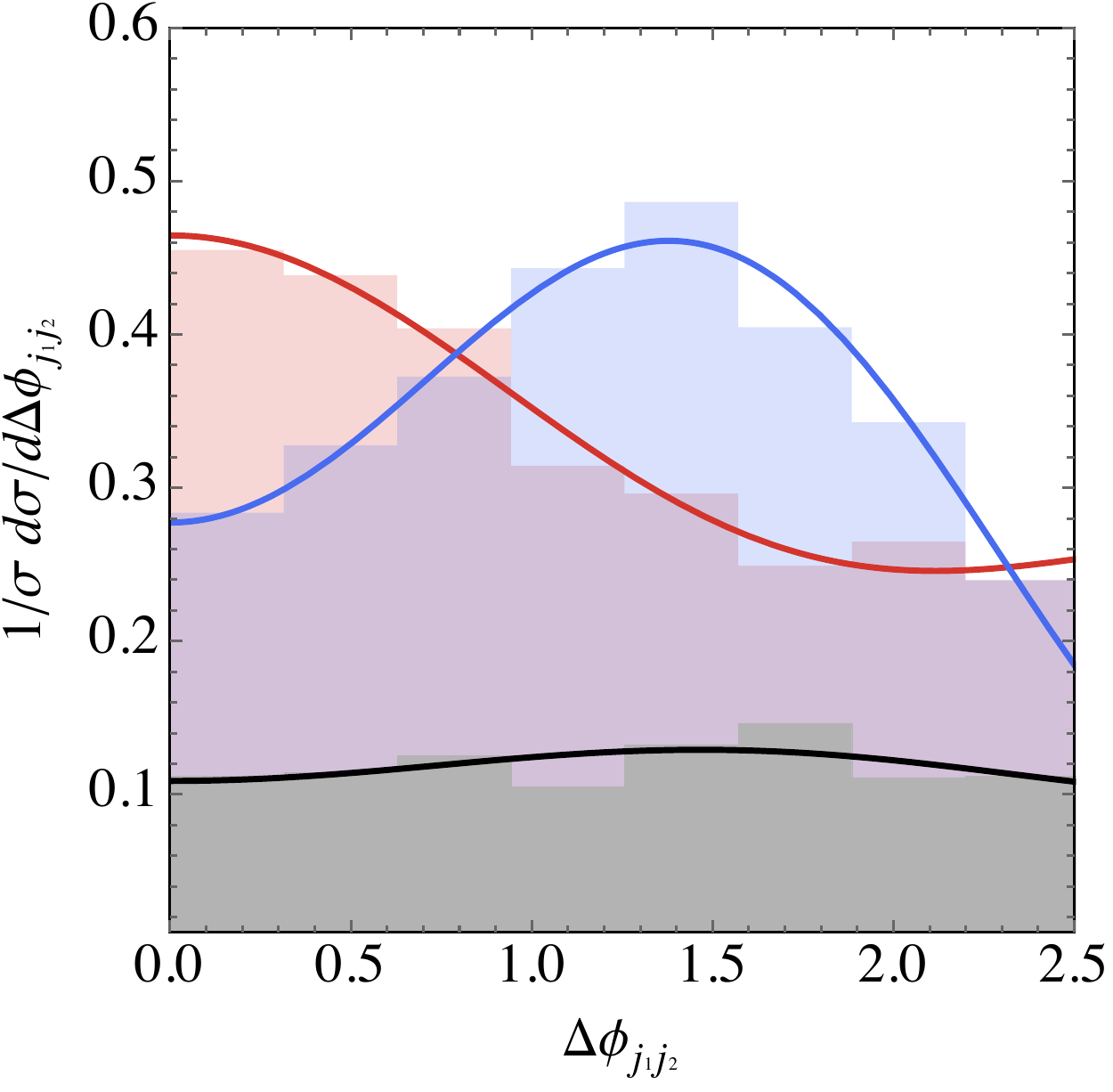} 

\vspace{2mm}

\includegraphics[width=0.45 \textwidth]{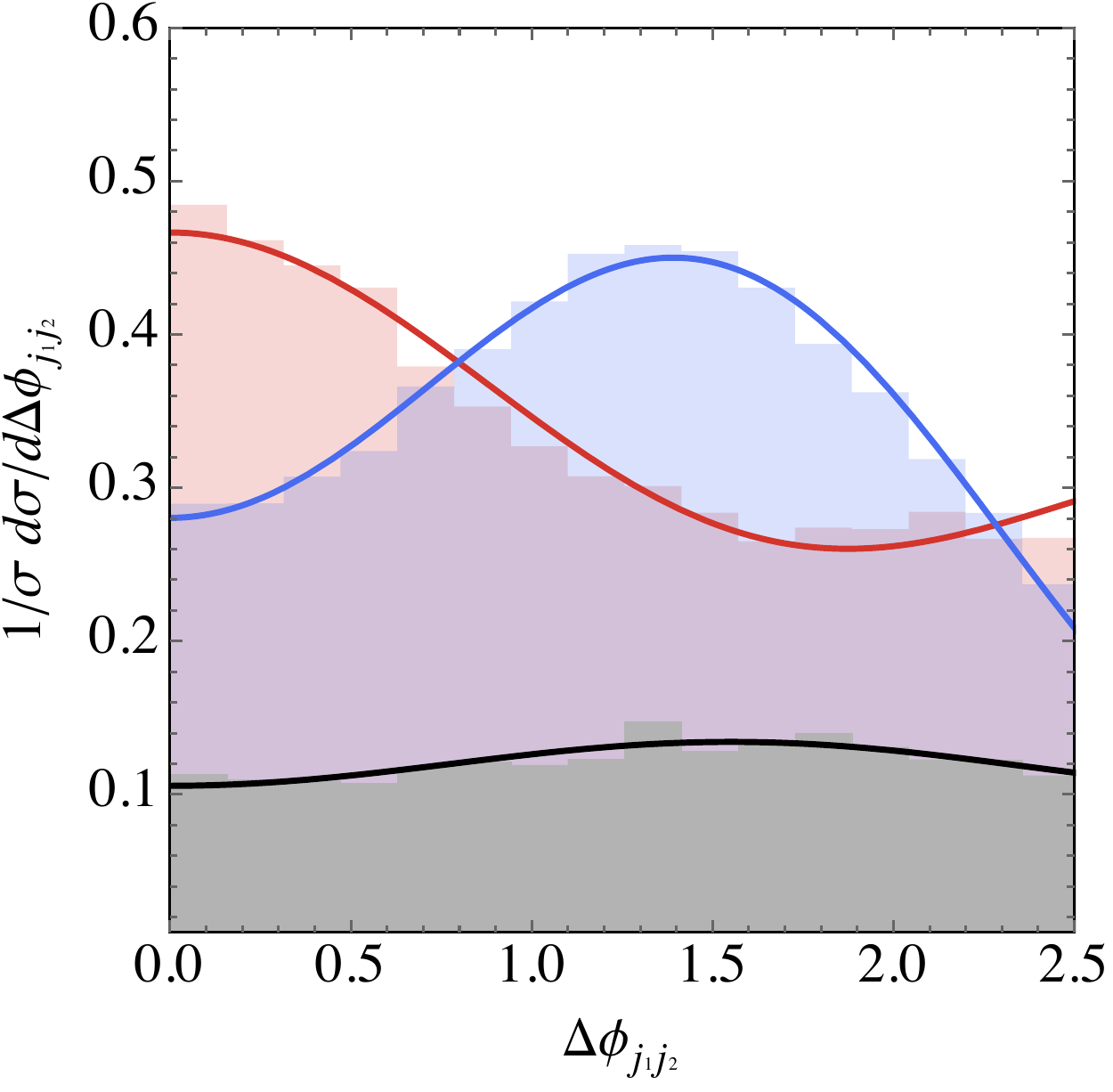} 
\caption{\label{fig:6}  Normalised $\Delta \phi_{j_1 j_2}$ distributions for $300 \, {\rm fb}^{-1}$ (upper panel) and $3000 \, {\rm fb}^{-1}$ (lower panel) of $14 \, {\rm TeV}$ LHC data. The red (blue) histogram shows the signal plus background prediction for $O_W$ ($O_{\tilde W}$). The grey bar chart represents the expected SM background, which for better visibility, has been  rescaled by a factor of $1/3$. The solid curves indicate the  best fits of the form $a_0 + a_1 \cos \Delta \phi_{j_1 j_2} +  a_2 \cos \left (2 \hspace{0.25mm} \Delta \phi_{j_1 j_2} \right)$. See text for additional explanations.}
\end{center}
\end{figure}

In  Fig.~\ref{fig:6} we present the results of our toy MC. The upper panel (lower panel) corresponds to $300 \, {\rm fb}^{-1}$ ($3000 \, {\rm fb}^{-1}$) of LHC data collected at $14 \, {\rm TeV}$. The expected azimuthal angle distributions for the signal plus background predictions are coloured blue (red) for $O_W$~($O_{\tilde W}$). For comparison, the SM-only  result (grey) divided by a factor of $3$ is also shown. The solid curves illustrate the best fits  to~(\ref{eq:20}), restricting the  rapidity separation $\Delta \phi_{j_1 j_2}$ to the range $[0, 2.5]$.   For $300 \, {\rm fb}^{-1}$ of data, we obtain   for $r_2$ the central values and uncertainties
\begin{equation}  \label{eq:21}
\begin{split}
&\left ( r_2 \right )_{W+ {\rm SM}}  = 0.15  \pm 0.10  \,, \\[1mm]
&\left ( r_2 \right )_{\tilde W + {\rm SM}}  = -0.45  \pm 0.14  \,, \\[1mm]
&\left ( r_2 \right )_{\rm SM}  = -0.12 \pm 0.22 \,.
\end{split}
\end{equation}
In the case of  $3000 \, {\rm fb}^{-1}$ of luminosity, we  find instead 
\begin{equation}  \label{eq:22}
\begin{split}
&\left ( r_2 \right )_{W+ {\rm SM}}  = 0.18  \pm 0.03 \,, \\[1mm]
&\left ( r_2 \right )_{\tilde W + {\rm SM}}  = -0.40  \pm 0.04  \,, \\[1mm]
&\left ( r_2 \right )_{\rm SM}  = -0.13 \pm 0.07 \,.
\end{split}
\end{equation}
We observe  that for $O_W$ ($O_{\tilde W}$) the combination  $r_2$  is indeed positive (negative). Defining a significance as $s_k = \big ( (r_2)_{k+{\rm SM}} - (r_2)_{\rm SM} \big)/(\Delta r_2)_{k + {\rm SM}}$, we get from~(\ref{eq:21}) the values $s_W = 2.7$ and $s_{\tilde W}=-2.4$, while~(\ref{eq:22}) leads to $s_W = 10.3$ and $s_{\tilde W}=-6.8$. Our  toy MC study corresponding to $300 \, {\rm fb}^{-1}$ ($3000 \, {\rm fb}^{-1}$) of data hence suggest that  a distinction between  the azimuthal angle distributions of $O_W$ and $O_{\tilde W}$ at the $5\sigma$ ($17 \sigma$) level should be possible at the $14 \, {\rm TeV}$ LHC. We emphasise that our toy study assumes a perfect detector and that we have not  optimised the cuts (\ref{eq:18}) to achieve the best significance. Once the data is on tape, it will become an experimental issue of how stringent the VBF-like selections can be made to extract the most information on the jet-jet angular correlations for a given limited sample size. 

\section{Conclusions}
\label{sec:4}

In this article we  have studied LHC constraints on effective dimension-7 operators that couple DM to the SM electroweak  gauge bosons and emphasised the complementarity of different $\slashed{E}_T$ searches for constraining the associated Wilson coefficients. Focusing on the interactions that induce only velocity-suppressed annihilation rates, we  have combined the information on all individual search modes that are available after LHC run-1. In this way we are able to derive bounds on the new-physics scale $\Lambda$ that exceed all previous limits. Our studies show that at present, depending on the choice of parameters, either mono-photon or mono-jet searches provide the most severe constraints on the considered dimension-7 interactions. For DM masses $m_\chi \lesssim 100 \ {\rm GeV}$ and Wilson coefficients $|C_k (\Lambda)| \simeq 1$, the existing $8 \, {\rm TeV}$ LHC searches allow to exclude values of $\Lambda$ below about $600 \, {\rm GeV}$ at $95 \% \, {\rm CL}$. The improved reach of $\slashed{E}_T$ analyses in 2015 and beyond is also studied, finding that with $25 \, {\rm fb}^{-1}$ of $14 \, {\rm TeV}$ data, LHC mono-jet searches should be able to improve the latter bound to approximately $1.3 \, {\rm TeV}$. Beyond this point  further progress will be hindered by the imperfect understanding of irreducible SM backgrounds such as $pp \to Z \, (\to \bar \nu \nu) + j$. Finding ways to overcome these limitations will be crucial to exploit the full physics potential of $\slashed{E}_T$ searches to be carried out at  later stages of the LHC. 

We have furthermore emphasised that given the large statistics expected at the phase-1 and phase-2 upgrades of the $14 \, {\rm TeV}$ LHC, $\slashed{E}_T$ searches should be able to not only determine integrated, but also differential cross sections. From the theoretical point of view, such normalised distributions have the clear advantage, that compared to the total cross sections theoretical uncertainties are reduced and that the obtained predictions depend only weakly on the assumptions underlying the EFT description. As an example we have explored the prospects to measure jet-jet angular correlations in $\slashed{E}_T + 2j$ events. Taking into account the pseudorapidity correlations of the two tagging jets, the resulting distributions in the azimuthal angle separation $\Delta \phi_{j_1 j_2}$ exhibit the relative strength of CP-even and CP-odd  interactions of DM with gauge boson pairs. Our toy MC studies indicate that already with $300 \, {\rm fb}^{-1}$ of data a distinction between the new-physics and the SM-only hypotheses  can be achieved at a statistically significant level, and that the sensitivity of the discussed searches is greatly improved by going to $3000 \, {\rm fb}^{-1}$ of luminosity.  A more precise determination of the analysing power, including systematic uncertainties would require a full detector simulation, which is beyond the scope of the present article. We however believe that it is imperative that the ATLAS and CMS collaborations  direct some activity towards the study of differential distributions of final states like $\slashed{E}_T + 2 j$. 

\begin{acknowledgments} 
We thank Tim Tait for clarifying discussions concerning his work \cite{Carpenter:2012rg} and are grateful to Benjamin~Fuks, Emanuele~Re and Giulia~Zanderighi for help with  MadGraph and/or PYTHIA. AC is supported by a Marie~Curie Intra-European Fellowship of the European Community's $7^{th}$ Framework Programme under contract number PIEF-GA-2012-326948. UH  acknowledges the hospitality and support of the CERN theory division and the Munich Institute for Astro- and Particle Physics (MIAPP) of the DFG cluster of excellence ``Origin and Structure of the Universe''. The research of AH is supported by an STFC Postgraduate Studentship. 
\end{acknowledgments}

\end{document}